\documentclass[conference]{IEEEtran}

\usepackage{cite}

\ifCLASSINFOpdf
  \usepackage[pdftex]{graphicx}
\else

  \usepackage[dvips]{graphicx}
  \DeclareGraphicsExtensions{.eps}
\fi

\usepackage[cmex10]{amsmath}
\usepackage{amssymb}
\usepackage{algorithm}
\usepackage{algorithmic}
\usepackage{bbm}
\usepackage{array}
\usepackage{epstopdf}
\usepackage{setspace}
\IEEEoverridecommandlockouts

\begin{document}

\title{Joint Pushing and Caching with a Finite Receiver Buffer: Optimal Policies and Throughput Analysis}

\begin{spacing}{0.973}

\author{\IEEEauthorrefmark{1}\IEEEauthorrefmark{2} \IEEEauthorblockN{Wei Chen}
\IEEEauthorblockA{\IEEEauthorrefmark{1} Department of Electronic Engineering / TNList\\
Tsinghua University\\
Beijing, China, 100084\\
Email: wchen@tsinghua.edu.cn, wchern@Princeton.EDU}
\and
\IEEEauthorrefmark{2} \IEEEauthorblockN{H. Vincent Poor}
\IEEEauthorblockA{\IEEEauthorrefmark{2} Department of Electrical Engineering\\
Princeton University\\
New Jersey, USA, 08544\\
Email: poor@Princeton.EDU}

\thanks{This research was supported in part by the US National Science Foundation under Grant CCF-1420575 and the NSFC Excellent Young Investigator Award under Grant No.61322111.}
}

\maketitle

\begin{abstract}
Pushing and caching hold the promise of significantly increasing the throughput of content-centric wireless networks. However, the throughput gain of these techniques is limited by the buffer size of the receiver. To overcome this, this paper presents a Joint Pushing and Caching (JPC) method that jointly determines the contents to be pushed to, and to be removed from, the receiver buffer in each timeslot. An offline and two online JPC policies are proposed respectively based on noncausal, statistical, and causal content Request Delay Information (RDI), which predicts a user's request time for certain content. It is shown that the effective throughput of JPC is increased with the receiver buffer size and the pushing channel capacity. Furthermore, the causal feedback of user requests is found to greatly enhance the performance of online JPC without inducing much signalling overhead in practice.
\end{abstract}

\IEEEpeerreviewmaketitle

\section{Introduction}
Mobile social applications have undergone explosive growth in the past decade, which has stimulated a dramatically increasing demand for bandwidth. Due to the scarcity of wireless spectrum, traditional radio access networks can hardly cope with the dramatically increasing data traffic, because the potential of increasing bandwidth, spatial reuse, and spectral efficiency has already been heavily exploited. Pushing and proactive caching, which hold the promise of providing substantial capacity gains by disseminating popular content in the idle spectrum when the network is off-peak, have been devised as an emerging and powerful solution that has attracted considerable attention recently.

The idea of adaptive pushing based on data popularity learning has been studied for over a decade \cite{Learn.Push}. More recently, the landmark works \cite{Wornell} and \cite{M.Ali} have revealed the fundamental limits of caching from an information-theoretic perspective. They motivated an extensive study of caching to exploit its multi-casting gain and off-peak access opportunities in 5G systems \cite{Leung}, \cite{Debbah}. It was shown in \cite{Sherman}, \cite{Caire}, and \cite{V.Lau} that caching may efficiently reduce the access latency, the outage probability, and the average transmit power in cloud-based mobile services, device-to-device (D2D) networks, and cooperative MIMO systems, respectively. An experimental study \cite{Experiment} demonstrated the practical performance of caching through prototype implementation. Furthermore, proactive caching was found to fully utilize not only the idle spectrum in off-peak times, but also the harvested renewable energy \cite{W.Chen}. However, pushing and caching must consume some storage resource for data buffering as a price to be paid for the spectral and energy efficiency gain. Many recent works, \cite{M.Ali}, \cite{V.Lau}, and \cite{H.Liu}, have noticed that the buffer size limits the performance of caching.

In this paper, we are interested in a joint pushing and caching system in which both the receiver buffer size and the pushing link capacity are limited. In this case, pushing content too much earlier than when it is requested may induce buffer overflow because too much content has to be cached in the buffer until they are read and removed. On the other hand, pushing content too late means missing user requests due to the limited pushing channel capacity. As a result, we should not only carefully schedule when and which content to be pushed, but also decide which content should be removed from the receiver buffer. Our aim is to maximize the effective throughput, which is defined as the number of content items read from the buffer when requested by a user.

To achieve this goal, we adopt content request delay, a random variable indicating when a particular content item is demanded by an individual user, to predict the user's requests. This is in contrast to previous works in which the request for content is predicted via its popularity over a group of users. The request delay information, also referred to as RDI, can be available in noncausal, statistical, or causal form. With noncausal RDI, an offline Joint Pushing and Caching (JPC) policy is proposed and analyzed based on network calculus, which characterizes the throughput upper bound of JPC. By noting that only statistical or causal RDI is available in practice, we present two online JPC policies based on trellis-aided dynamic programming, which maximizes the expected throughput in low complexity. Both theoretical analysis and simulation results reveal that the throughput of JPC is increased with the receiver buffer size and the pushing channel capacity. Furthermore, causal feedback of RDI is found to achieve a significant throughput gain, especially when the buffer size is relatively small.

\section{System Model}

Consider a wireless link from a Base Station (BS) to a user. The BS has $L$ content files, $W_i$, $i=1, \ldots, L$, each consisting of $B$ bits and being generated at the origin of the timeline, i.e., $t=0$. The user asks for a content item after a random delay since its generation at $t=0$. Without loss of generality, we assume that a content item will not be requested twice. In this case, we can use a nonnegative random variable $X_i$ to denote the request delay for content item $W_i$. If the user never wants to read $W_i$, then $X_i = \infty$. Let us define the content request delay vector as $\mathbf X = [x_1, \ldots, x_L]$, the probability density function ($p.d.f.$) of which is denoted by $p_{\mathbf X}(\mathbf x)$. We assume that the $X_i$ are independent, i.e., $p_{\mathbf X}(\mathbf x)= \prod_{i=1}^{L} p_{X_i}(x_i)$, where $p_{X_i}(x_i)$ stands for the marginal $p.d.f.$ of $X_i$. The content items have a lifetime given by $t^{\max} = \min \{t : p_{X_i}(x_i) \equiv 0, \forall i, \forall t \leq x < \infty\}$. After the lifetime, none of the content items will be required. A finite lifetime is assumed, i.e., $t^{\max}<\infty$.

In contrast to conventional communication systems, where the BS transmits a content item $W_i$ only if the user asks for it, JPC allows the BS to transmit content items to the user before they are requested. In order to cache this pushed content, the user is equipped with a buffer that is capable of caching at most $NB$ bits, or equivalently, $N$ content files. For convenience, we shall also refer to $N$ as the buffer size.

\begin{figure}[!t]
\centering
\includegraphics[width=3.5in]{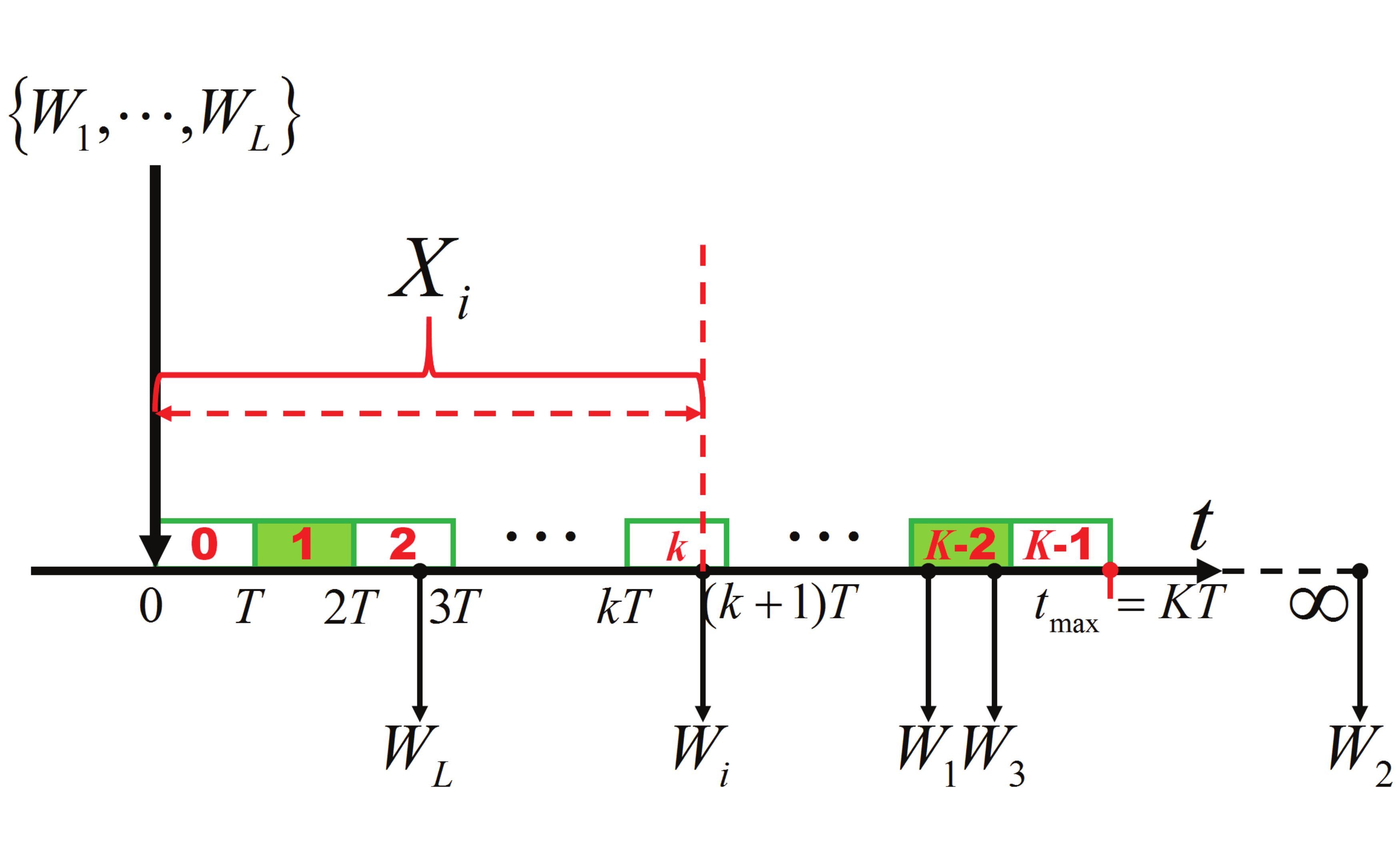}
\caption{Content Request Delay. In this realization, content items $W_i$ and $W_L$ are requested in the $k$th and second timeslots, respectively. Both $W_1$ and $W_3$ are requested in the $(K-2)$th timeslots. The user will not ask for content item $W_2$.}
\label{RequestDelay}
\end{figure}

When the BS pushes a content item, the transmission rate is $C$ bits/s. As a result, it takes $T=\frac{B}{C}$ seconds to push one content item to the user's buffer.\footnote{A careful reader may notice that the rate of a wireless link can be time-varying due to fading or opportunistic spectrum access. However, it should also be noted that a content file is usually large, and hence it needs a sufficiently long time to be sent. Because the channel is approximately ergodic in this case, the transmission time and the average rate can be considered to be deterministic.} The time consumed in pushing one content file is referred to as a timeslot. Therefore, by the $k$th timeslot, we mean the time duration of $[kT,(k+1)T)$. Since the content items have a finite lifetime, a content item demanded by the user should be requested before the $K$th timeslot, where $K = \lceil \frac{t^{\max}}{T} \rceil$, as shown in Fig. \ref{RequestDelay}.

We refer to the above model as an $(N, K, p_{\mathbf X})$ joint pushing and caching system, or simply, $(N, K, p_{\mathbf X})$-pushing. Let us introduce the following notation that will be used later. The set of content items requested in the $k$th timeslot is denoted by $\mathcal W^u[k]$, i.e., $\mathcal W^u[k] = \{W_i:  kT \leq X_i < (k+1)T\}$. Let $q_{i,k}$ denote the probability that content item $W_i$ is requested in the $k$th timeslot. From the $p.d.f.$ of the content request delay $X_i$, we know that $q_{i,k} = \int_{kT}^{(k+1)T} p_{X_i}(x) \textrm{d}x$. If the content request delays are independent and identically distributed ($i.i.d.$), all content items will have the same possibility of being requested in the $k$th timeslot, i.e., $q_{i,k} = q_k$, for all $i$.

To characterize a JPC policy, we let $W[k]$ and $\mathcal W^r[k]$ denote the content item pushed by the BS in the $k$th timeslot and the set of content items removed from the receiver buffer at the end of the $k$th timeslot, respectively. If the BS pushes nothing in the $k$th timeslot, we let $W[k]=\emptyset$. In this context, $W[k]$ and $\mathcal W^r[k]$ jointly determine the set of content items cached in the buffer at the beginning of the $k$th timeslot, which we shall refer to as buffer state $\mathcal S_k$. More specifically, we have\footnote{For two sets $\mathcal A$ and $\mathcal B$, $\mathcal A \setminus \mathcal B := \{a: a \in \mathcal A \textrm{ and } a \not\in \mathcal B\}$.}
\begin{equation}
\label{buffer_state_update}
\mathcal S_{k+1} = \mathcal S_k \cup \{W[k]\} \setminus \mathcal W^r[k].
\end{equation}
Note that the buffer states $\mathcal S_k$ has three constraints: Its initial state is an empty set, i.e., $\mathcal S_k = \emptyset$; its cardinality is less than or equal to the buffer size, i.e., $|\mathcal S_k| \leq N$; and it has bounded time variation given by $ | \mathcal S_k \setminus \mathcal S_{k-1} | \leq 1$, because the BS can transmit at most one content item to the user in a single timeslot.

For the sake of Quality of Service (QoS) assurance, if a content item cannot be found in the buffer when the user asks for it, it should be transmitted over a traditional link. In other words, the more content items that are in the buffer when requested, the more idle spectrum in off-peak time is effectively utilized. As a result, we define the effective throughput of $(N, K, p_{\mathbf X})$-pushing to be the average number of content items that can be read from the buffer when they are requested. It can be written as
\begin{equation}
\label{throughput_def}
R= \sum_{i=1}^{L} \Pr \left\{ W_i \in \mathcal S_{\lfloor \frac{X_i}{T} \rfloor} \right\}.
\end{equation}
In other words, $(N, K, p_{\mathbf X})$-pushing provides $RB$ effective bits that are desired by the user. Hence, the throughput $R$ can be regarded as a key performance metric of $(N, K, p_{\mathbf X})$-pushing. Because the throughput $R$ is determined by the buffer states $\mathcal S_k$, our purpose is to design JPC policies that maximize the throughput $R$ by adjusting the buffer states $\mathcal S_k$ appropriately, under three different assumptions on the availability of RDI.

\section{$(N, K, p_{\mathbf X})$-pushing with Noncausal RDI}

In this section, we propose and analyze an offline JPC under the assumption that the content request delays are known when the content items are generated at $t=0$.

\subsection{Optimal Policy}

Having noncausal request delay information, the BS knows which content items will be requested in the $k$th timeslot. In this case, we may maximize the throughput by having as many content items as possible, which will be requested in the $k$th timeslot, cached in the user's buffer at the beginning of this timeslot. To achieve this goal, we propose a greedy but optimal policy, which pushes content to the user unless the receiver buffer is full. The content file pushed in the $k$th timeslot is chosen according to the following two rules. First, the content file is not already cached in the user's buffer. Second, the content file will be requested in the nearest future after the $k$th timeslot.\footnote{A content item requested in the $k$th timeslot should not be pushed in this timeslot because it will no longer be needed when it is completely transmitted at timepoint $(k+1)T$.}

At the end of the $k$th timeslot, the receiver buffer will remove the content items that have been read by the user in the $k$th timeslot. As a result, the removed content items in the $k$th timeslot belong to the intersection of the set of requested content items, $\mathcal W^u[k]$, and the buffer state $\mathcal S_k$ of this timeslot.

Based on the above, the optimal offline JPC policy can be characterized by $W[k]$ and $\mathcal W^r[k]$ as
\begin{equation}
\label{noncausal_push}
W[k] = \left\{
\begin{array}{cl}
\underset{ \{W_i:  W_i \not\in \mathcal S_k \textrm{ and } X_i \geq (k+1)T\} }{\arg \min} X_i & \textrm{ if } S_k < N \\
\emptyset & \textrm{ if } S_k = N,
\end{array}
\right.
\end{equation}
and
\begin{equation}
\label{noncausal_remove}
\mathcal W^r[k] = \mathcal W^u[k] \cap \mathcal S_k,
\end{equation}
where the buffer state $\mathcal S_k$ is updated according to Eq. (\ref{buffer_state_update}).

\subsection{Throughput Analysis}

For $i.i.d.$ request delays, we are capable of analyzing the throughput of $(N, K, p_{\mathbf X})$-pushing with noncausal RDI. Let $s_k$ and $m_k$ respectively denote the number of content items cached in the buffer, and the number of content items requested by the user in the $k$th timeslot, i.e., $s_k = |\mathcal S_k|$ and $m_k = |\mathcal W^u[k]|$, where $|\cdot|$ represents the cardinality of a set, and the initial value of $s_k$ is $s_0=0$. From network calculus theory \cite{Cruz}, the update equation of $s_k$ can be written as
\begin{equation}
\label{update_cached_number}
s_{k+1} = \left[ \left( s_k - m_k \right)^{+} + 1 \right] \wedge N,
\end{equation}
where $(x)^+ := \max \{x, 0\}$ and $x \wedge y := \min \{x, y\}$.

Let us denote the request pattern vector by $\mathbf m = \left[ m_0, m_1, \ldots, m_{K-1} \right]$. For any given $\mathbf m$, $s_k$ can be calculated from Eq. (\ref{update_cached_number}). Furthermore, we know that $m_k \wedge s_k$ content items are read from the buffer in the $k$th timeslot. Because $\mathbf m$ follows a multinomial distribution with probability mass function $p(\mathbf m)=\frac{L!}{\prod_{k=0}^{K-1} m_k !}\prod_{k=0}^{K-1} q_k ^{m_k}$, the throughput of the offline JPC policy with noncausal RDI can be obtained by
\begin{equation}
\label{noncausal_iid_throughput}
R = L! \sum_{||\mathbf m||_1 = L} \frac{\prod_{k=0}^{K-1} q_k ^{m_k}}{\prod_{k=0}^{K-1} m_k !} \sum_{k=1}^{K-1} m_k \wedge s_k,
\end{equation}
where $||\cdot||_1$ denotes the $1$-norm of a vector.

Eqs. (\ref{update_cached_number}) and (\ref{noncausal_iid_throughput}) present a computational approach to calculating the throughput of offline JPC. However, its complexity can be high for large $K$ and $L$. To give further insight, we present an approximate throughput analysis for the scenario in which the request delays have identical uniform distributions, i.e., $X_i \sim U(0,KT)$ for all $i$. The approximate analysis relies on the observation that despite how the content requests are distributed in $N$ consecutive timeslots, the user can read at most $N$ content files from its buffer due to the buffer size constraint. Since the probability that a content item is requested in the $N$ consecutive timeslots is given by $\frac{N}{K}$, we have\footnote{Due to space limitation, we omit the proof of Eq. (\ref{noncausal_iid_uniform}) but simply sketch its key idea in this paragraph.}
\begin{equation}
\label{noncausal_iid_uniform}
\begin{split}
R \approx & \left( 1 - \frac{N}{K} \right)L + \left(1 - \frac{K}{N}\right) \times \\
& \sum_{j=1}^{L-N} j {{L}\choose{N+j}} \left( \frac{N}{K} \right)^{N+j} \left( 1 - \frac{N}{K} \right)^{L-N-j}.
\end{split}
\end{equation}
When $K$ is large, Eq. (\ref{noncausal_iid_uniform}) can be further approximated by \cite{Table}
\begin{equation}
\label{noncausal_iid_uniform_apx}
R \approx NK \left(1-e^{-\frac{L}{NK}}\right),
\end{equation}
which implies that $R \to L$ when $K \to\infty$. Note that the total number of timeslots, $K$, increases linearly with the capacity of the pushing channel from the BS to the user. If the pushing channel capacity is sufficiently high, almost all the content items can be read from the buffer when they are requested.

\section{$(N, K, p_{\mathbf X})$-pushing with Statistical RDI}

In this section, we are interested in the scenario in which only statistical information about $\mathbf X$, namely, its $p.d.f.$, $p_{\mathbf X}(\mathbf x)$, is available. To obtain the optimal online JPC policy based on statistical RDI, we formulate an optimization problem relying on $p_{\mathbf X}(\mathbf x)$ only. Note that a content item contributes to the throughput, if and only if this item is in the receiver buffer when it is requested. As a result, Eq. (\ref{throughput_def}) can be rewritten as $R = \sum_{i=1}^{L} \sum_{k=1}^{K} q_{i,k} \mathbbm 1 \left\{ W_i \in S_k \right\}$, where $\mathbbm 1 \left\{ \cdot \right\}$ is an indicator function. By interchanging the order of summation and noting that $\sum_{i=1}^{L}q_{i,k} \mathbbm 1 \left\{ W_i \in S_k \right\} = \sum_{W_i \in \mathcal S_k} q_{i,k}$, we have $R = \sum_{k=1}^{K} \sum_{W_i \in \mathcal S_k} q_{i,k}$. Also recalling the three constraints on $\mathcal S_k$, we formulate an optimization problem given by
\begin{equation}
\label{stat_opt}
\begin{array}{rrcl}
\displaystyle \max_{\mathcal S_k} & \multicolumn{3}{l}{ \sum_{k=1}^{K-1} \sum_{W_i \in \mathcal S_k} q_{i,k} } \\
\textrm{s.t.} & |\mathcal S_k \setminus \mathcal S_{k-1}| & \leq & 1  \\
& |\mathcal S_k| & \leq & N \\
& \mathcal S_0 & = & \emptyset,
\end{array}
\end{equation}
where the variables are buffer states $\mathcal S_k$ for $k=1, \ldots, K-1$.

\subsection{Optimal Policy}

\begin{figure}[!t]
\centering
\includegraphics[width=3.01in]{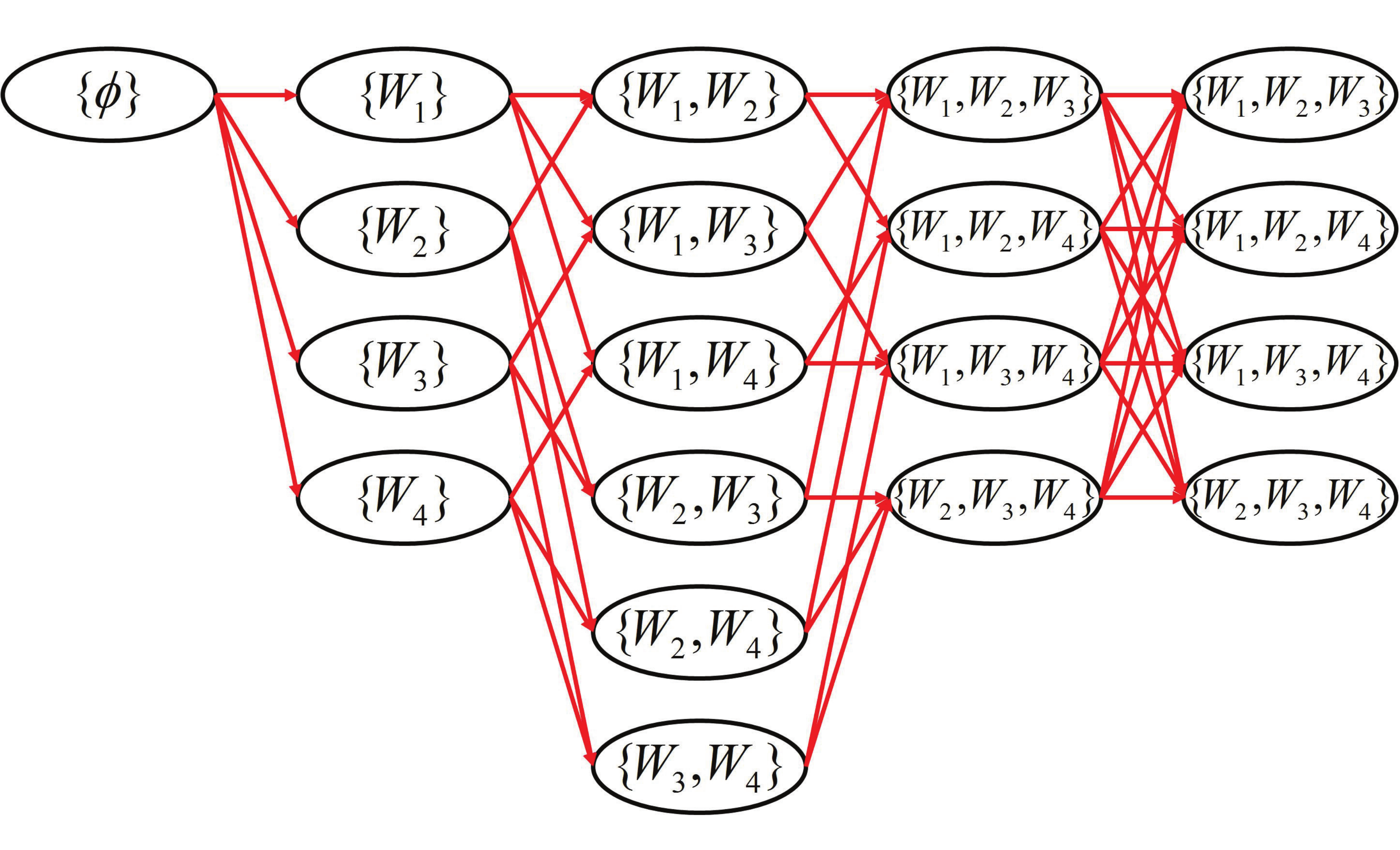}
\caption{Trellis Graph that Characterizes the Legitimate Transitions of Buffer States. In this trellis graph, we set $L=4$, $N=3$, and $K=5$.}
\label{TrellisGraph}
\end{figure}

To solve problem (\ref{stat_opt}) with low complexity, we use dynamic programming \cite{Bellman}. Let us first characterize the legitimate transitions of buffer states by a trellis graph, as shown in Fig. \ref{TrellisGraph}. The trellis graph has $K$ vertical slices. The $k$th vertical slice consists of all possible buffer states in the $k$th timeslot. More specifically, since the buffer caches $k \wedge N \wedge L$ content items in the $k$th timeslot\footnote{If $\mathcal S'_k$ is a strict subset of $\mathcal S_k$, then $\sum_{W_i \in \mathcal S'_k} q_{i,k}$ is strictly less than $\sum_{W_i \in \mathcal S_k} q_{i,k}$. As a result, we should have as many content items as possible cached in the buffer, in order to increase the probability that the user finds its desired content from its buffer. To achieve this goal, the BS should keep pushing content items to the user until its buffer is full, and then always keep the receiver buffer full.}
and there are a total of $L$ different content items, the $k$th vertical slice consists of ${L}\choose{k \wedge N \wedge L}$ different buffer states. To distinguish different buffer states in the same vertical slice, we assign another subscript $r$ chosen from an integer set $\left\{1, \ldots, {{L}\choose{ k \wedge N \wedge L}} \right\}$ to each buffer state in the $k$th timeslot, thereby rewriting it as $\mathcal S_{k,r}$.\footnote{There are multiple methods to assign the subscripts. For instance, we first map each $\mathcal S_k$ to a number $\varpi(\mathcal S_k) = \sum_{i=1}^{L} 2^i \mathbbm 1 \{W_i \in \mathcal S_k\}$. For given $k$, buffer state $\mathcal S_k$ with the $r$th smallest $\varpi(\mathcal S_k)$ is assigned subscript $r$.}

In the trellis graph, two buffer states, $\mathcal S_{k,r_1}$ and $\mathcal S_{k+1,r_2}$, which belong to two neighbor vertical slices respectively, are connected by an edge if and only if they have at most one different content item, i.e., $|\mathcal S_{k+1,r_2} \setminus \mathcal S_{k,r_1}| \leq 1$. The consecutive edges from $\mathcal S_0$ to any buffer state in the $(K-1)$th vertical slice form a complete path. Each complete path represents a feasible solution to (\ref{stat_opt}). If a path passes through a buffer state $\mathcal S_{k,r}$, it receives a reward of $\sum_{i \in \mathcal S_{k,r}} q_{i,k}$. Our aim is to find the optimal path that receives the maximal total reward.

To achieve this goal, we present an iterative algorithm as follows. Let us define the survival path of buffer state $\mathcal S_{k,r}$ to be the path from $\mathcal S_0$ to $\mathcal S_{k,r}$, which has the maximal total reward. We characterize the survival path of $\mathcal S_{k,r}$ by a set-valued vector $\mathbf V(\mathcal S_{k,r}) = \left[\mathcal S_0, \mathcal S_{1,r_1(\mathcal S_{k,r})}, \ldots, \mathcal S_{k-1,r_{k-1}(\mathcal S_{k,r})}, \mathcal S_{k,r}\right]$, where $\mathcal S_{j,r_j(\mathcal S_{k,r})}$ denotes the $j$th buffer state in the survival path of $\mathcal S_{k,r}$. Furthermore, we let $\Gamma(\mathcal S_{k,r})$ denote the total reward of the survival path of $\mathcal S_{k,r}$. Given all the survival paths to the buffer states in the $k$th vertical slice, we may obtain the survival path of buffer state $\mathcal S_{k+1,r}$ as
\begin{equation}
\label{path_update}
\mathbf V(\mathcal S_{k+1,r}) = \left[\mathbf V(\mathcal S_{k,r^*(\mathcal S_{k+1,r})}), \mathcal S_{k+1,r}\right],
\end{equation}
where
\begin{equation}
\label{path_select}
\mathcal S_{k,r^*(\mathcal S_{k+1,r})} = \underset{|\mathcal S_{k+1,r} \setminus \mathcal S_{k,u}| \leq 1 }{\arg \max} \Gamma(\mathcal S_{k,u}).
\end{equation}
The new survival path $\mathbf V(\mathcal S_{k+1,r})$ has an updated reward given by
\begin{equation}
\label{reward_update}
\Gamma(\mathcal S_{k+1,r}) = \Gamma(\mathcal S_{k,r^*(\mathcal S_{k+1,r})}) + \sum_{W_i \in \mathcal S_{k+1,r}} q_{i,k}.
\end{equation}

By implementing Eqs. (\ref{path_update}), (\ref{path_select}), and (\ref{reward_update}) iteratively, we obtain ${{L}\choose{ {(K-1)} \wedge N \wedge L}} $ survival paths, each ending at a different buffer state in the $(K-1)$th vertical slice. Finally, we select the optimal complete path having the maximal total reward as
\begin{equation}
\label{opt_path}
\mathbf V^* = \underset{ \mathcal S_{K-1,r} }{\arg \max} \quad \Gamma(\mathcal S_{K-1,r}).
\end{equation}

Let us rewrite the optimal complete path to be $\mathbf V^* = \left[\mathcal S_0, \mathcal S^*_1, \ldots, \mathcal S^*_{K-1} \right]$, the elements of which represent the optimal solution to problem (\ref{stat_opt}). From the optimal buffer states, the content items to be pushed by the BS and removed from the receiver buffer in the $k$th timeslot are obtained by
\begin{equation}
\label{stat_push}
W[k] = \mathcal S^*_{k+1} \setminus \mathcal S^*_k,
\end{equation}
and
\begin{equation}
\label{stat_remove}
\mathcal W^r[k] = \mathcal S^*_k \setminus \mathcal S^*_{k+1},
\end{equation}
which characterize the optimal JPC with only statistical RDI.

\subsection{Throughput Analysis}

Since the throughput of $(N, K, p_{\mathbf X})$-pushing with statistical RDI is presented by the objective function of problem (\ref{stat_opt}), the maximal throughput can be obtained by
\begin{equation}
\label{stat_throughput}
R =  \sum_{k=1}^{K-1} \sum_{W_i \in \mathcal S^*_k} q_{i,k},
\end{equation}
which is also equivalent to $\underset{ \mathcal S_{K-1,r} }{\max} \quad \Gamma(\mathcal S_{K-1,r})$.

For $i.i.d.$ request delays, we are capable of presenting an analytical result for the throughput. By noting that all the buffer states of the $k$th vertical slice have the same reward $(k \wedge N \wedge L)q_k$, we have
\begin{equation}
\label{stat_iid}
R = \sum_{k=1}^{N \wedge L-1} k q_k + (N \wedge L) \sum_{k= N \wedge L}^{K-1} q_k,
\end{equation}
which is upper bounded by $N \wedge L$.

When all the request delays are uniformly distributed, i.e., $X_i \sim U(0,KT)$, we may substitute $q_k = \frac{1}{K}$, for $0 \leq k \leq K-1$, into Eq. (\ref{stat_iid}). This yields
\begin{equation}
\label{stat_iid_uniform}
R = (N \wedge L) \left( 1- \frac{N \wedge L+3}{2K} \right),
\end{equation}
which approaches $N \wedge L$ for large $K$. By comparing Eqs. (\ref{noncausal_iid_uniform_apx}) and (\ref{stat_iid_uniform}), we notice that in contrast to the offline JPC that can benefit from the noncausal RDI, the throughput of online JPC with statistical RDI is always bounded by the buffer size, regardless of the pushing link capacity. To overcome this, we shall optimize the online JPC by exploiting causal feedback.

\section{$(N, K, p_{\mathbf X})$-pushing with Causal RDI}

In this section, we study online $(N, K, p_{\mathbf X})$-pushing, where causal feedback of the user's content requests is enabled. By causal feedback, we mean that the user tells the BS at the end of the $k$th timeslot which content items are requested in this timeslot. Such feedback will not induce much signalling overhead, because only the indices of the requested content items are transmitted. More specifically, at most $L \log L$ bits are transmitted due to the feedback of $L$ content requests. Based on the assumption that a content item will not be requested twice, causal feedback may effectively avoid wasting timeslots or buffer space on pushing or caching outdated content items that will not be needed by the user again.

\subsection{Optimal Policy}

Optimal $(N, K, p_{\mathbf X})$-pushing with causal RDI relies on a method of formulating and solving optimization problems iteratively. This is due to the fact that the set of outdated content items is updated at the end of each timeslot. As a result, a new optimization problem should be formulated to maximize the expected throughput in the remaining timeslots.

From the causal feedback, the BS knows at time $kT$ the set of requested content items in the $(k-1)$th timeslot, namely, $\mathcal W^u[k-1] = \{W_i: (k-1)T \leq X_i < kT\}$. Hence, it can update the outdated content set as
\begin{equation}
\label{outdated_content}
\mathcal W^o[k] = \bigcup_{j=1}^{k} \{W_i: (j-1)T \leq X_i < jT\}.
\end{equation}

Let $\mathcal S_k^{\sharp}$ denote the buffer state in which the content items in $\mathcal W^u[k-1]$ have been removed from the buffer. Before showing how $\mathcal S_k^{\sharp}$ is updated, we first formulate the $k$th timeslot's optimization problem based on the outdated content set $\mathcal W^o[k]$ and the current buffer state $\mathcal S_k^{\sharp}$ as
\begin{equation}
\label{causal_opt}
\begin{array}{rrcl}
\displaystyle \max_{\mathcal S_j, j \geq k+1} & \multicolumn{3}{l}{\sum_{j=k+1}^{K-1} \sum_{W_i \in \mathcal S_j} \frac{q_{i,j}}{\sum_{l=k+1}^{K-1}q_{i,l}} } \\
\textrm{s.t.} & |\mathcal S_{j+1} \setminus \mathcal S_j| & \leq & 1  \\
& |\mathcal S_{k+1} \setminus \mathcal S_k^{\sharp}| & \leq & 1  \\
& |\mathcal S_j| & \leq & N \\
& \mathcal S_j \cap \mathcal W^o[k] & = & \emptyset.
\end{array}
\end{equation}
In problem (\ref{causal_opt}), the last constraint means that none of the buffer states in the remaining timeslots should consist of any outdated content. Since both $\mathcal S_0^{\sharp}$ and $\mathcal W^o[0]$ are empty sets, problem (\ref{causal_opt}) reduces to problem (\ref{stat_opt}) in the $0$th timeslot.

Having established the optimization problem for the $k$th timeslot, we turn our attention to how it is solved and updated. Similar to (\ref{stat_opt}), problem (\ref{causal_opt}) can be solved by implementing iterations (\ref{path_update})-(\ref{opt_path}). There are only two differences to be noted in the trellis graph of problem (\ref{causal_opt}). First, the original buffer state is $\mathcal S_k^{\sharp}$ rather than $\mathcal S_0$. Second, the buffer states in the trellis graph do not contain any outdated content in $\mathcal W^o[k]$.

The optimal solution to (\ref{causal_opt}) gives the optimal buffer state $\mathcal S^*_{k+1}$ before removing the content items in $\mathcal W^u[k]$. Therefore, the buffer state at the beginning of the $(k+1)$th timeslot is updated to be
\begin{equation}
\label{remove_requested_content}
\mathcal S_{k+1}^{\sharp} = \mathcal S^*_{k+1} \setminus \mathcal W^u[k],
\end{equation}
from which we formulate the optimization problem for the $(k+1)$th timeslot.

By formulating and solving problem (\ref{causal_opt}) iteratively, we obtain $\mathcal S^*_k$ and $\mathcal S_k^{\sharp}$ for all $k$. As a result, the content item to be pushed by the BS in the $k$th timeslot is obtained by
\begin{equation}
\label{causal_push}
W[k] = \mathcal S^*_{k+1} \setminus \mathcal S_k^{\sharp}.
\end{equation}
By noting that not only the content items that are not in the estimated optimal buffer state $\mathcal S^*_{k+1}$, but also the content items requested in the $k$th timeslot should be removed from the receiver buffer at the end of the $k$th timeslot, we have
\begin{equation}
\label{causal_remove}
\mathcal W^r[k] = ( \mathcal S_k^{\sharp} \setminus \mathcal S^*_{k+1} ) \cup (\mathcal W^u[k] \cap \mathcal S^*_{k+1}).
\end{equation}

\subsection{Throughput Analysis}

We study the performance of online JPC with causal feedback under the assumption of $i.i.d.$ content request delays. Similar to subsection III-B, a computational approach based on enumerating $\mathbf m$ is presented first. In the $k$th timeslot, there are $L- \sum_{j=0}^{k-1}m_j$ content items that are not outdated. Because $s_k$ of them are cached, the probability that a requested content item can be found in the buffer is obtained by $\frac{s_k}{L- \sum_{j=0}^{k-1}m_j} \wedge 1$. Since $m_k$ content items are requested in the $k$th timeslot, the throughput of $(N, K, p_{\mathbf X})$-pushing with causal RDI can be written as
\begin{equation}
\label{causal_iid_throughput}
R = L! \sum_{||\mathbf m||_1 = L} \frac{\prod_{k=0}^{K-1} q_k ^{m_k}}{\prod_{k=0}^{K-1} m_k !}\sum_{k=1}^{K-1} \frac{m_k s_k}{L- \sum_{j=0}^{k-1}m_j} \wedge m_k ,
\end{equation}
where $s_k$ is given by Eq. (\ref{update_cached_number}). As with Eq. (\ref{noncausal_iid_throughput}), the computational complexity of Eq. (\ref{causal_iid_throughput}) can be high for large $K$ and $L$.

When all the request delays are independently and uniformly distributed, we may present an approximate but analytical result for the throughput for large $K$, which is given by\footnote{Due to space limitations, we omit the proof of Eqs. (\ref{causal_iid_uniform}) and (\ref{causal_iid_uniform_apx}).}
\begin{equation}
\begin{split}
\label{causal_iid_uniform}
R \approx {K \choose L}^{-1} \sum_{i=1}^{L} \sum_{k=0}^{K-L} & \frac{i \wedge k \wedge N}{i} {{L+k-i-1} \choose {L-i}} \\
& \times {{K-L+i-k} \choose {i-1}}.
\end{split}
\end{equation}

For large $N$ and $L$, Eq. (\ref{causal_iid_uniform}) can be further simplified to be
\begin{equation}
\label{causal_iid_uniform_apx}
R \approx (N \wedge L)\left( \ln \frac{L}{N \wedge L} + 1 \right).
\end{equation}
Eq. (\ref{causal_iid_uniform_apx}) implies that the throughput of $(N, K, p_{\mathbf X})$-pushing with causal RDI cannot approach $L$ unless the buffer size is large enough to cache all the content items, i.e. $N \geq L$.

\section{Simulation Results}

In this section, numerical results are presented to validate the theoretical analysis and demonstrate the potential of the JPC policies. Throughout this section, we assume that there are totally $L=10$ content items. The content request delays are $i.i.d.$ random variables obeying uniform distributions.

\begin{figure}[!t]
\centering
\includegraphics[width=3.4in]{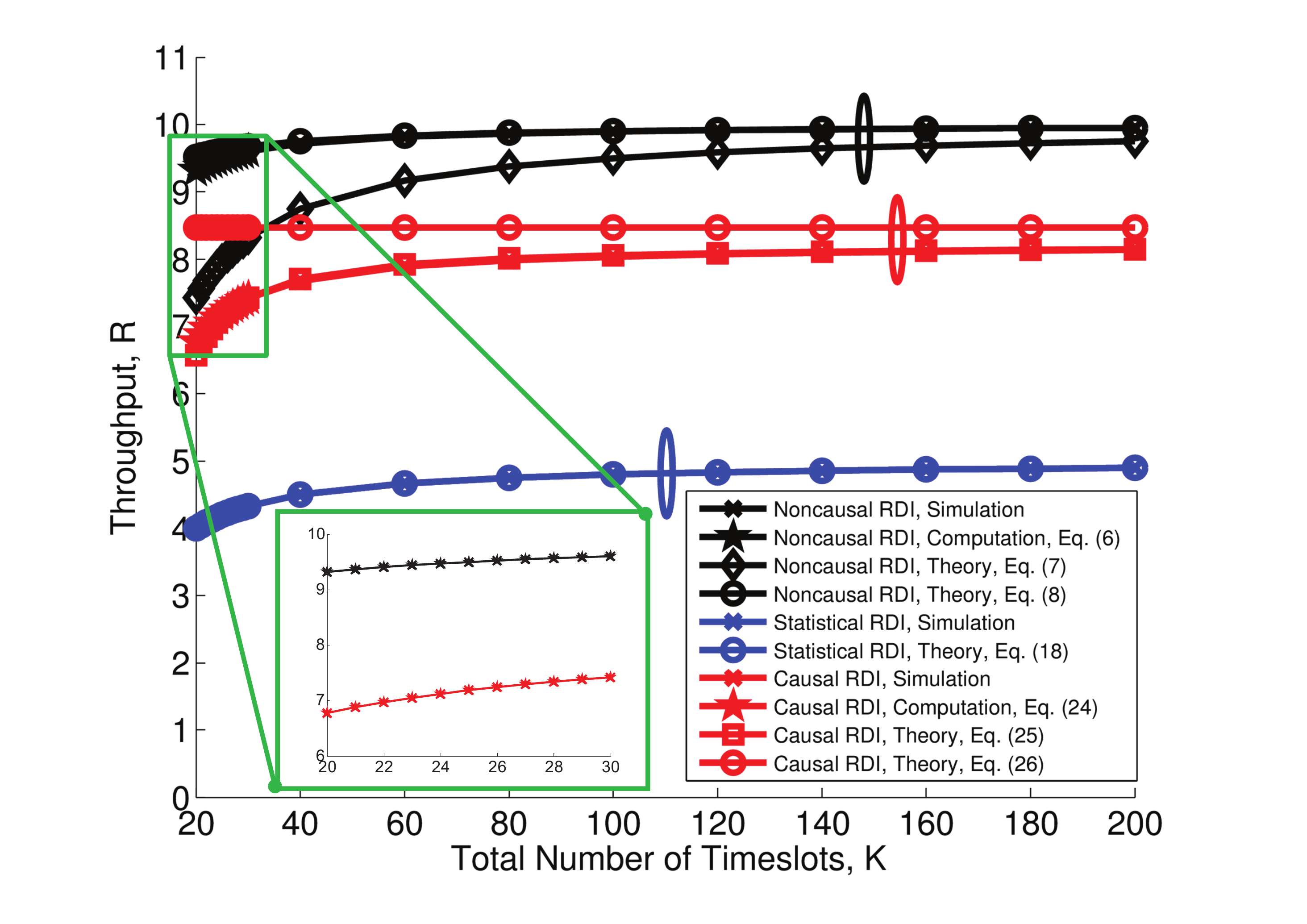}
\caption{Throughput versus Total Number of Timeslots}
\label{curve_RvsK}
\end{figure}

Fig. \ref{curve_RvsK} presents the throughput $R$ versus total number of timeslots $K$, where the buffer size is set to be $N=5$. It is observed that the throughput obtained via Eq. (\ref{stat_iid_uniform}), as well as computations using Eqs. (\ref{noncausal_iid_throughput}) and (\ref{causal_iid_throughput}) perfectly match their corresponding simulation results.\footnote{Due to the computational complexity, we cannot adopt Eqs. (\ref{noncausal_iid_throughput}) and (\ref{causal_iid_throughput}) to calculate the throughput when $K$ is large. Hence, we present computational results only for $K \le 30$, as shown by the zoomed-in curves in Fig. \ref{curve_RvsK}.} The approximate but analytical throughput results given by Eqs. (\ref{noncausal_iid_uniform}), (\ref{noncausal_iid_uniform_apx}), (\ref{causal_iid_uniform}), and (\ref{causal_iid_uniform_apx}) approach the simulation results asymptotically for large $K$. In particular, the approximation errors are upper bounded by $10 \%$, when $K$ is greater than $60$.

Having validated the theoretical analysis, we turn our attention to the comparison of the three JPC policies when $K$ is large. The offline JPC policy achieves a throughput of approximately $10$ thanks to its knowledge of noncausal RDI. In other words, all the content items can be read from the buffer with high probabilities when they are requested. However, noncausal RDI is not available in practice. In this case, online JPC with only statistical RDI suffers from a throughput loss of $50\%$. Fortunately, causal feedback of the user's requests recovers over $60 \%$ of the throughput loss without making any impractical assumption on RDI. Furthermore, it is worth noting that $K$ increases linearly with the pushing channel capacity. As a result, Fig. \ref{curve_RvsK} also reveals how the throughput increases with the pushing channel capacity.

\begin{figure}[!t]
\centering
\includegraphics[width=3.19in]{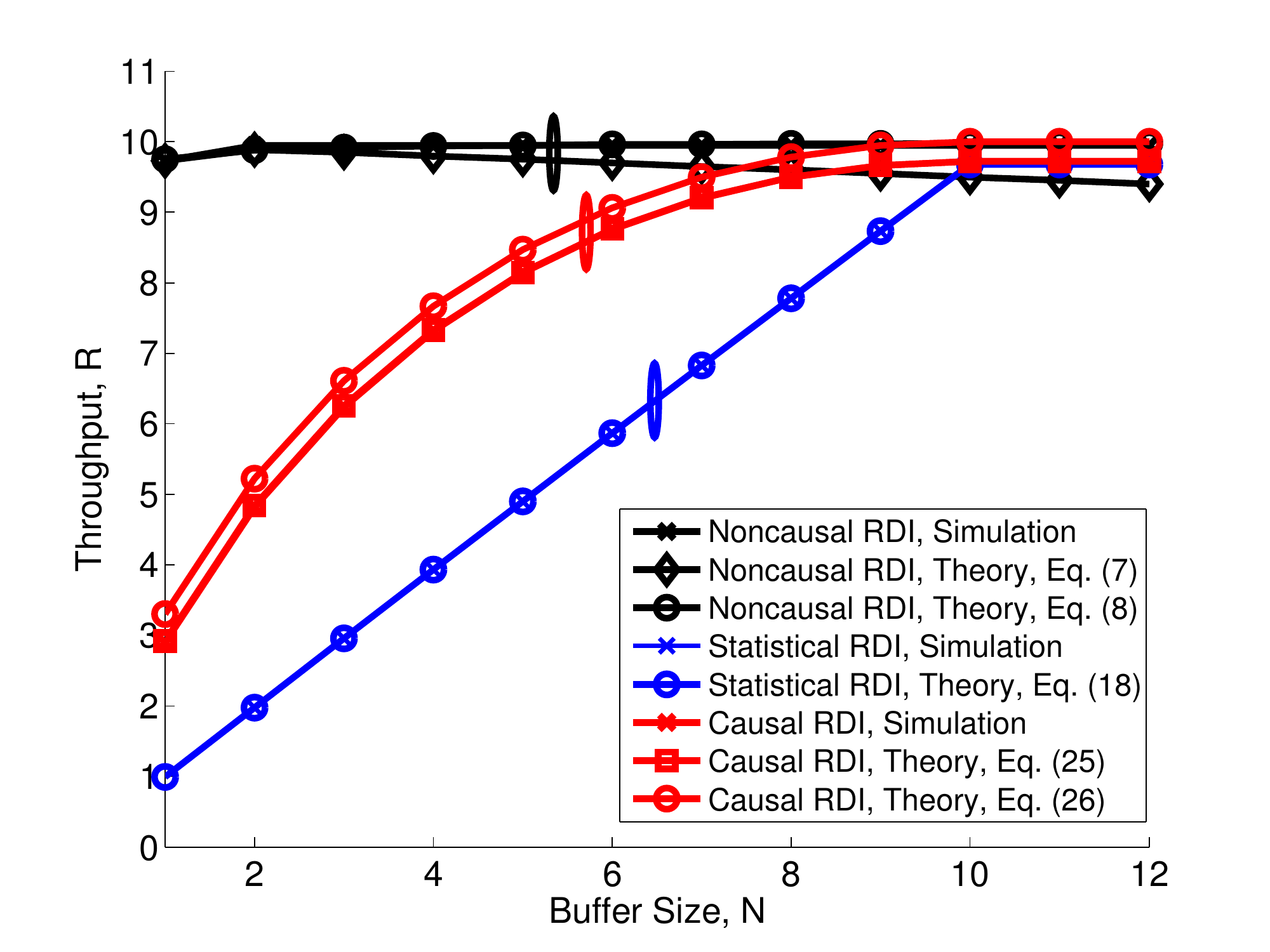}
\caption{Throughput versus Buffer Size}
\label{curve_RvsN}
\end{figure}

Fig. \ref{curve_RvsN} presents the throughput $R$ versus buffer size $N$, where the total number of timeslots is set to be $K=200$. Again, the theoretical results given by Eqs. (\ref{noncausal_iid_uniform}), (\ref{noncausal_iid_uniform_apx}), (\ref{stat_iid_uniform}), (\ref{causal_iid_uniform}), and (\ref{causal_iid_uniform_apx}) match their corresponding simulation results very well. It is not surprising that the throughput of all the JPC policies increases with the buffer size. For offline JPC, an increase in buffer size brings marginal throughput gain because the throughput has closely approached its upper bound $L=10$ even when $N=1$. In contrast, the buffer size dominates the performance of online JPC policies. When only statistical RDI is available, the throughput increases linearly with the buffer size, thereby being very limited in the small buffer case. Fortunately, the causal feedback of RDI can bring over $100\%$ throughput gain to online JPC when the buffer size is less than $4$. This comparison demonstrates the potential of feedback in practice. Therefore, it is worth paying a small cost of feedback overhead in order to achieve the corresponding significant throughput gain.

\section{Conclusion}

In this paper, we have studied $(N, K, p_{\mathbf X})$-pushing with a finite receiver buffer in three typical scenarios, in which noncausal, statistical, and causal request delay information is available, respectively. With noncausal RDI, an offline JPC has been presented to reveal a throughput upper bound of $(N, K, p_{\mathbf X})$-pushing. With statistical and causal RDI, two online JPC schemes that adopt trellis-aided dynamic programming to maximize the expected throughput have been proposed from a more practical perspective. Both theoretical analysis and simulation results show that the throughput of $(N, K, p_{\mathbf X})$-pushing increases with the receiver buffer size and the pushing channel capacity. Furthermore, since the causal feedback of content requests may significantly improve the throughput of online JPC, feedback mechanisms are of great importance in practice, particularly when the buffer size is relatively small.

\end{spacing}

\end{document}